\documentclass[10pt]{article}
\usepackage[utf8]{inputenc}
\usepackage[english]{babel}
\usepackage{amssymb}
\usepackage{bm}
\usepackage{amsmath}
\usepackage{natbib}
\usepackage{graphicx}
\usepackage{geometry}
\usepackage{textcomp}
\usepackage{subfigure}
\usepackage{xcolor}
\usepackage{ulem}

\title{A Linear Approximation for the Effect of Cylindrical Differential Rotation on Gravitational Moments: Application to the Non-Unique Interpretation of Saturn's Gravity}
\author{Yayaati Chachan and David J. Stevenson \\
Division of Geological and Planetary Sciences, California Institute of Technology \\
1200 E California Blvd, Pasadena, USA, 91125}

\begin{document}
\maketitle

\begin{abstract}
The higher order gravitational moments of a differentially rotating planet are greatly affected or even dominated by the near surface differential flow. Unlike the contribution from rigid body rotation, this part of the gravity field can be well estimated by linear theory when the differential rotation is on cylinders and the corresponding gravity field arises from the higher order moments directly introduced by that flow. In the context of an n = 1 polytrope, we derive approximate analytical formulas for the gravity moments. We find that $\Delta J_{2\mathrm{n}}$ is typically at most a few times $(-1)^{n+1} \; a \; q \; d^{5/2}$, where $a$ is the amplitude of the differential rotation (as a fraction of the background rigid body rotation), $q = \Omega^2 R^3 / G M$ is the usual dimensionless measure of rotation for the planet (mass $M$, radius $R$), and $d << 1$ is the characteristic depth of the flow as a fraction of the planetary radius. Applied to Saturn, with $a$ set by the observed surface wind amplitude, we find first that the observed signs of the $\Delta J_{2\mathrm{n}}$ are a trivial consequence of the definition of the corresponding Legendre polynomials, but the {\it Cassini} observations can not be explained by a simple exponentially decaying flow and instead require a substantial retrograde flow at depth and a larger $d$ than the simple scaling suggests. This is consistent with the results reported by \citet{Iess2019}. However, there is no fluid dynamical requirement that the flows observed in the atmosphere are a guide to the flows thousands of km deeper. We explore a wide range of flow depths and amplitudes which yield values for $\Delta J_{\mathrm{n}}$ that are acceptable within the error estimates and thus highlight the inherent non-uniqueness of inferences made from the higher order gravity moments.
\end{abstract}

\section{Introduction} \label{sec:intro}

The external gravity field of an isolated, uniformly rotating, hydrostatic planet is expressible in the form

\begin{equation}
V_{ext} (r, \theta) = \frac{G M}{r} \left[ 1 - \sum_{n=1}^{\infty} \bigg(\frac{a}{r}\bigg)^{2n}  J_{2n} P_{2n}(cos \theta) \right]
\label{eq:V_ext}
\end{equation}

where $J_n$, the gravitational moments are given by:

\begin{equation}
J_n = - \frac{1}{M a^n} \int_{R^3} \rho (\bm{r}) r^n P_n (cos \theta) \bm{\mathrm{d}^3r}
\label{eq:moment}
\end{equation}

where $M$ is planet mass, $a$ is planet radius (equatorial or volume averaged), $P_n$ is the order $n$ Legendre polynomial, $\rho$ is the density, $G$ is the gravitational constant, and $\theta$ is the colatitude. The dimensionless coefficients of this expansion are called gravitational moments and denoted $J_{2n}$ where $n$ is a positive integer. They exist because the planet rotates. The expansion contains only even moments because the rotational perturbation is quadratic in rotation rate and therefore does not distinguish between Northern and Southern hemispheres. In linear response theory, $J_2$ is proportional to $q = \Omega^2 R^3 / G M$ and all higher $J$'s are zero, where $\Omega$ is the rotation rate, $R$ is the planet's radius, $M$ is the planet's mass, and $G$ is the gravitational constant. This assumes rigid body rotation. However, $q$ is not very small for giant planets and so the linear response is far from adequate, especially given the high accuracy with which the gravity field can be measured by orbiting spacecraft such as {\it Juno} and {\it Cassini}. The higher $J$'s arise explicitly from non-linearity; that is, the centrifugal potential is exactly of degree 2, but the change in shape and resulting self gravity introduces the higher order terms so that $J_{2n}$ is approximately proportional to $q^n$ \citep{Hubbard1975,Hubbard1982}. Moreover, even low order moments are affected by higher order corrections. Although the $J$'s are linearly related to the density anomalies of the same harmonic degree, the underlying non-linearity makes the theory complicated, even for simple models of planetary structure, and it can be difficult to develop an intuition for the dependence of $J$'s on features of that structure, which is, after all, the whole point of making the precise measurements. However, a great deal of successful effort has been put into forward modeling of the $J$'s from physical models of the planet's interior \citep[e.g.][]{Hubbard2013,Galanti2017,Cao2017a}. This is necessarily non-unique (you cannot invert the $J$'s to determine planet structure except in the context of models with only a few degrees of freedom) but it has provided us with a great deal of understanding of fluid planets.

However, fluid bodies do not rotate as rigid bodies. This is self-evident in the zonal flows observed in their atmospheres. In general, this flow need not be purely cylindrical, a special case of the Taylor-Proudman theorem (i.e. with no variation of the East-West flow along an axis parallel to the rotation axis). Deviations from the Taylor-Proudman theorem in a fluid with large density variations but no MHD effects arise primarily from latitudinal entropy gradients but also potentially from the Reynolds stress \citep[e.g., Equation 22 of][]{Liu2008,Kaspi2009}. The advantage of assuming cylindrical differential rotation is that it allows for the definition of a centrifugal potential, the gradient of which provides the acceleration and thus the wind strength at each cylindrical radius. At first sight, it might seem that even this simple kind of differential rotation only exacerbates the difficulty of developing an intuitive or approximate understanding of the $J$'s. However, there is a sense in which the effect of differential rotation may actually be easier to understand than the part due to rigid body rotation. The reason is that the centrifugal potential is no longer just degree 2, but contains higher order moments, much higher if the differential rotation is largely confined to a thin shell. One can then compute the gravitational response to that part alone using linear theory. Moreover, we can then approximate the planet as spherical. This will be an adequate approximation (with a fractional error of order $q$) provided the differential rotation is sufficiently small relative to the total rotation. This is the approach presented in this paper. It assumes that one can isolate the contribution to the $J$'s that arises from the differential rotation from the often larger or comparable part that arises from the rigid part of the rotation. This can often be done because the higher order $J$'s arising from rigid body rotation are usually confined to a narrow range of possible values based on interior models that fit the lower $J$'s \footnote{This is predicated on the standard view that we actually know the structure of the outermost ten percent or so of the planet very well: an adiabatic hydrogen-dominated region for Jupiter and Saturn.} \citep[e.g.][]{Iess2019}. Moreover, the emphasis on great precision that usually accompanies analysis of the interior models need no longer apply -- one might well be content with getting the differential rotation of a planet to just tens of percent, especially if the inference is necessarily non-unique, but also because the observational data can have substantial errors at the higher order moments.

This paper is motivated by a desire to develop a better intuition about what it is that determines the differential rotation contribution to the $J$'s, more specifically how the amplitude and patterns of their contribution to the $J$'s is set by the amplitude, depth, and shape of the differential flow profile. It focuses on the high order $J$'s where the effect of differential rotation is more readily discerned in the data.

In keeping with our emphasis on transparency and simplicity, we use an n = 1 polytropic model (pressure everywhere proportional to the square of the density). This is not essential to the idea of our approach but it leads to a linear equation for hydrostatic equilibrium and enables the derivation of analytical formulas. As we shall explain, it is not necessary that the polytrope apply everywhere; models with a separate but small core can also be described in this way. The biggest error in this assumption lies in the breakdown of that equation of state as one approaches the surface, but this is still a smaller error than the error arising from assuming sphericity.

\begin{figure}
\centering
\includegraphics[width=\linewidth]{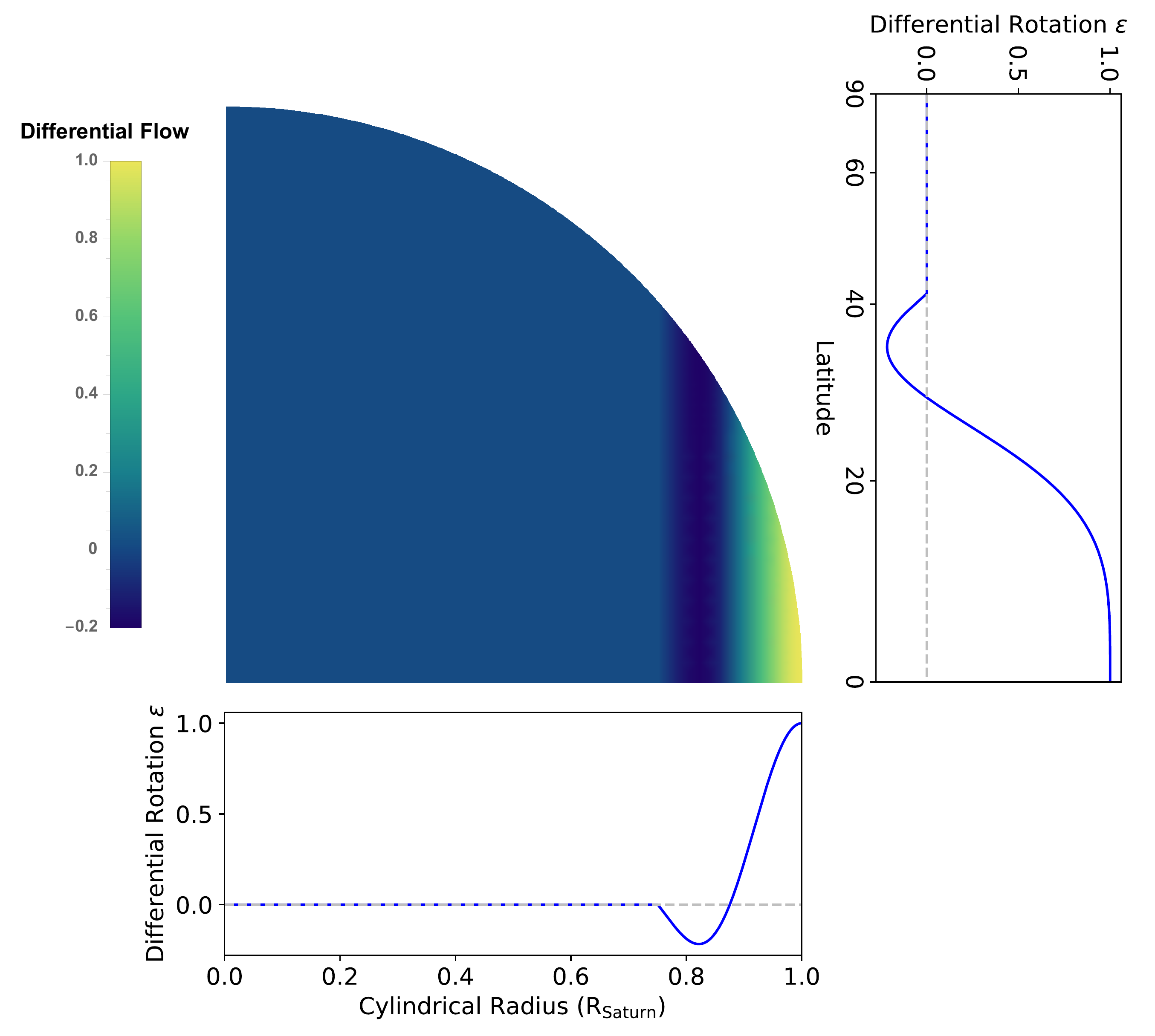}
\caption{A depiction of the cylindrical differential rotation in the meridional plane. 1D projections of the flow speed as a function of the cylindrical radius and latitude are also shown and they are clearly coupled. A deeper flow would therefore extend to higher latitudes and a shallower flow would be constrained to lower latitudes. In this work, we use the terms `deep' and `shallow' to refer to the {\it cylindrical} depth of the flow.}
\label{fig:model_sketch}
\end{figure}

In Section 2, we describe our model and derive the main results for an arbitrary choice of functional form for the cylindrical differential rotation. In Section 3, we show how these results can yield simple scaling laws and simple analytical formulas for the case of differential flow that is exponentially declining with depth. We also show how to estimate the corrections that arise from planetary oblateness (i.e. corrections of order $q$). In Section 4, we apply our procedure to the {\it Cassini} observations of Saturn and show how we can recover flows similar to those suggested by more precise calculation, but also how they can be quite non-unique. We conclude with a discussion of what this approach tells us about the general features of gravitational effects caused by differential rotation.

\section{Theoretical Model} \label{sec:theory}

If the angular velocity within the planet depends only on the cylindrical radius $r_c = r \sin \theta$, then the sole dependence of flow properties on $r_c$ allows us to write their effect on the hydrostatic equilibrium of the planet as a gradient of a potential $V_{rot}$:

\begin{equation}
\frac{1}{\rho} \nabla P = \nabla (V + V_{rot})
\label{eq:hydrostatic}
\end{equation}

where $\rho$ is the density, $P$ is the pressure, and $V$ is the gravitational potential. Figure~\ref{fig:model_sketch} shows an example of the differential flows considered in this work. All references to the depth of a flow apply to the {\it cylindrical} radius. We need an equation of state to relate the pressure to the density. This hydrostatic equation is difficult to solve for an arbitrary equation of state but it has a particularly simple form for one with polytropic index $n = 1$, i.e., $P = K \rho^2$. Here, $K$ is a constant of proportionality that is mainly determined by the properties of the degenerate electron gas but is also somewhat affected by the planet's entropy. Using this polytropic equation, the hydrostatic equilibrium simplifies to: 

\begin{equation}
\nabla (2 K \rho -V - V_{rot}) = 0
\end{equation}

This can only be satisfied if $2 K \rho = V + V_{rot}$. We assume that the contribution of differential rotation to the rotational potential is small compared to the background solid-body rotation potential. We normalize the potential to the monopole, $G M / R$, so we can write the rotational potential as:

\begin{equation}
V_{rot} = q \; \int_{0}^{s} (1 + \epsilon(s'))^2 s' \mathrm{d}s' \\
\approx q \; \int_{0}^{s} (1 + 2\epsilon(s')) s' \mathrm{d}s'
\label{eq:Vrot_integrate}
\end{equation}
where $s = r_c / R$ is the cylindrical radius normalised to the planet's radius, $\epsilon (s)$ is the differential flow profile as a function of $s$, the integrand is Taylor-expanded to first order under the assumption that $\epsilon << 1 $, and $q$ is given by

\begin{equation}
q = \Omega^2 R^3 / G M 
\end{equation}
Here, $G$ is the gravitational constant, $M$ is the planet's mass, $R$ its radius, and $\Omega$ is the background solid body rotation. The first term is simply the solid body rotation term. The second term, linear in $\epsilon$, is henceforth denoted $\delta V_{rot}$. The Poisson equation relates the Laplacian of the potential to the density and using the above relationship between density and potential, we obtain:

\begin{equation}
\nabla^2 \delta V = - 4 \pi G \rho = - k^2 (\delta V + \delta V_{rot})
\end{equation}
where $k^2$ = (2$\pi G) / K$, $\delta V$ is only the self gravity potential arising from $\delta V_{rot}$. We can solve this by writing $\delta V = V_h + V_p$ as a sum of a general (homogeneous) and a particular solution. The general solution is the solution to the homogeneous Helmholtz equation and is given by:

\begin{equation}
V_h = \sum_{Even \; n}  A_n \; j_n(kr) \; P_n(cos \theta) 
\label{eq:homogeneous}
\end{equation}
where $j_n$ are the spherical Bessel's functions, $P_n$ are Legendre Polynomials of order $n$, and $A_n$ are the calculable coefficients. To calculate the particular solution $V_p$, we make an additional assumption that the differential rotation is significant near the surface of the planet and that it decays away rapidly inside the planet. This is a reasonable assumption for Jupiter, perhaps less so for Saturn. In this approximation, the curvature terms in the cylindrical Laplacian can be ignored:

\begin{equation}
\frac{\mathrm{d}^2 V_p}{\mathrm{d} r_c^2} = - k^2 \delta V_{rot}
\label{eq:V_p_diffeq}
\end{equation}

Solving for $V_p$, we get an expression which is a function of cylindrical radius $r_c$. However, the homogeneous solution ($V_h$) and the external potential ($V_{ext}$ obtained from solving $\nabla^2 V_{ext} =0$) are normally expressed as a function of the spherical coordinates. We therefore need to project V$_p$ from cylindrical to spherical coordinates to match the boundary condition for the potential at the planet's surface. This can be done using the Legendre polynomials which form a complete basis set for axisymmetric functions of the spherical coordinates:

\begin{eqnarray}
V_p (r sin\theta) = \sum_{Even \; n} \; V_{pn}(r) \; P_n (cos\theta) \label{eq:V_p_projection1} \\
V_{pn}(r) = \frac{2n+1}{2} \int_{-1}^{1} V_p(r sin\theta) \; P_n(cos\theta) \; \mathrm{d}(cos\theta)
\label{eq:V_p_projection2}
\end{eqnarray}

We want to calculate the deviation in gravitational moments due to a small shallow flow and we can do that by applying boundary conditions to the potential at the planetary surface. We refer to these as $\Delta J_n$ where the $J$'s are defined by Equation~\ref{eq:moment}. We assume that this surface can be approximated by a sphere (we deal in Section~\ref{sec:oblate} with the error introduced by this). The potential and its derivative must be continuous at the surface which gives us two equations that relate the quantities $V_{pn}$, $A_n j_n (kr)$, and $\Delta J_n$ at $r = R$ and their derivatives. 

\begin{eqnarray}
V_{pn} (R) + A_n j_n (kR) = - \Delta J_n \\
R V_{pn}' + k R A_n j_n'(kR) = (n+1) \Delta J_n
\label{eq:boundary_cond}
\end{eqnarray}
Here, the prime $'$ refers to differentiation with respect to $r$ (so $j_n'(kR) = {\rm d} j_n(kr)/{\rm d}r$ evaluated at $r = R$). Eliminating $A_n$, we obtain the following expression for $\Delta J_n$:

\begin{equation}
\Delta J_n = \frac{R V'_{pn}(R) j_n(kR) - kR V_{pn}(R) j'_n(kR)}{(n+1)j_n(kR) + kR j'_n(kR)}
\label{eq:deltaj_n}
\end{equation}

For a planet with no core, $k R = \pi$. For a planet with a high density (rock or ice) core, $k R = \pi (1-\delta)$ where $\delta$ is the ratio of core to planet mass. However, this correction is small for the models we consider for the $\Delta J_n$'s. In the next section, we demonstrate the utility and power of this model that arises out of its simplicity.

\section{Analytical Solutions} \label{sec:analytic}
We use the theoretical framework established in the previous section to do analytic calculations for the gravitational moments due to differential flow in the limit of very shallow flow. Although, we do not use this analytic solution directly to fit Saturn's higher order gravitational moments, it is still instructive to perform this exercise to gain an understanding of how the deviations in gravitational moments relate to the properties of differential flow.

\subsection{Limit of Shallow Flow}
Let us assume that the differential flow profile can be represented by a steeply declining exponential which is only a function of the cylindrical radius of a fluid planet:

\begin{equation}
\epsilon (s) = a \; e^{-b \; (1-s)}
\end{equation}
Here, $a$ and $b$ are constants that characterize the amplitude and the depth of the flow, and $s = r_c / R$ is again the cylindrical radius as a fraction of the equatorial radius. We can integrate this function analytically to obtain $V_{rot}$ and calculate $V_p$ from the differential equation~\ref{eq:V_p_diffeq}, with the boundary condition that $V_p$ and its derivative vanish at $s = 0$. To proceed further with the analytic calculation, we need to project $V_p$ onto spherical coordinates using the Legendre Polynomials (equation~\ref{eq:V_p_projection1}). For very shallow flow, we can use a Taylor series expansion of the Legendre Polynomials up to a certain order (that we choose) and calculate the integrals analytically. We choose to expand the even Legendre Polynomials to second order in $\mathrm{cos} \; \theta$ around $\theta = \pi/2$:

\begin{equation}
P_n (\mathrm{cos} \; \theta) = {n \choose n/2} \frac{(-1)^{n/2} }{2^n} \bigg[ 1 - \frac{n \; (n+1)}{2} (\mathrm{cos} \theta)^2 \bigg]
\end{equation}
where $n \choose n/2$ is the binomial coefficient. With this approximation, we can project $V_p$ onto a basis set of Legendre polynomials with $s = r \; \mathrm{sin} \theta / R$. We then have an expression for $V_{pn}$ from equation~\ref{eq:V_p_projection2}. Given that we calculate this integral in the approximation of shallow flow, we can neglect all terms that do not contain the large exponential term in the expression for $V_p$. To calculate $V_{pn}$, we resort to the saddle point method (Appendix A) which works well for strongly peaked functions such as the exponential in the expression for $V_p$. Evaluating $V_{pn}$ and its derivative at $r = R = 1$ (fractional radius in our calculations) and substituting it in our expression for $\Delta J_n$, we obtain (see Appendix B for a detailed derivation):

\begin{multline}
\Delta J_n = - \frac{2n + 1}{2} {n \choose n/2} \frac {2 a  \; q \; \pi^2} {b^3} \frac{(-1)^{n/2} }{2^n} \sqrt{\frac{2 \pi}{b}}  \; \times \\
\frac{\bigg[ \left( -\frac{5}{2} + b + \frac{3}{2b} \right) + \frac{n (n+1)}{4b} \left( \frac{7}{2} -b - \frac{9}{2b} \right) \bigg] j_n(\pi) - \pi \bigg[ \left(1 - \frac{3}{b} \right) \left( 1 - \frac{n(n+1)}{4b} \right) \bigg] j'_n(\pi)}{(n+1)j_n(\pi) + \pi j'_n(\pi)}
\label{eq:O1border}
\end{multline}

\begin{figure}
\centering
\includegraphics[width=1\linewidth]{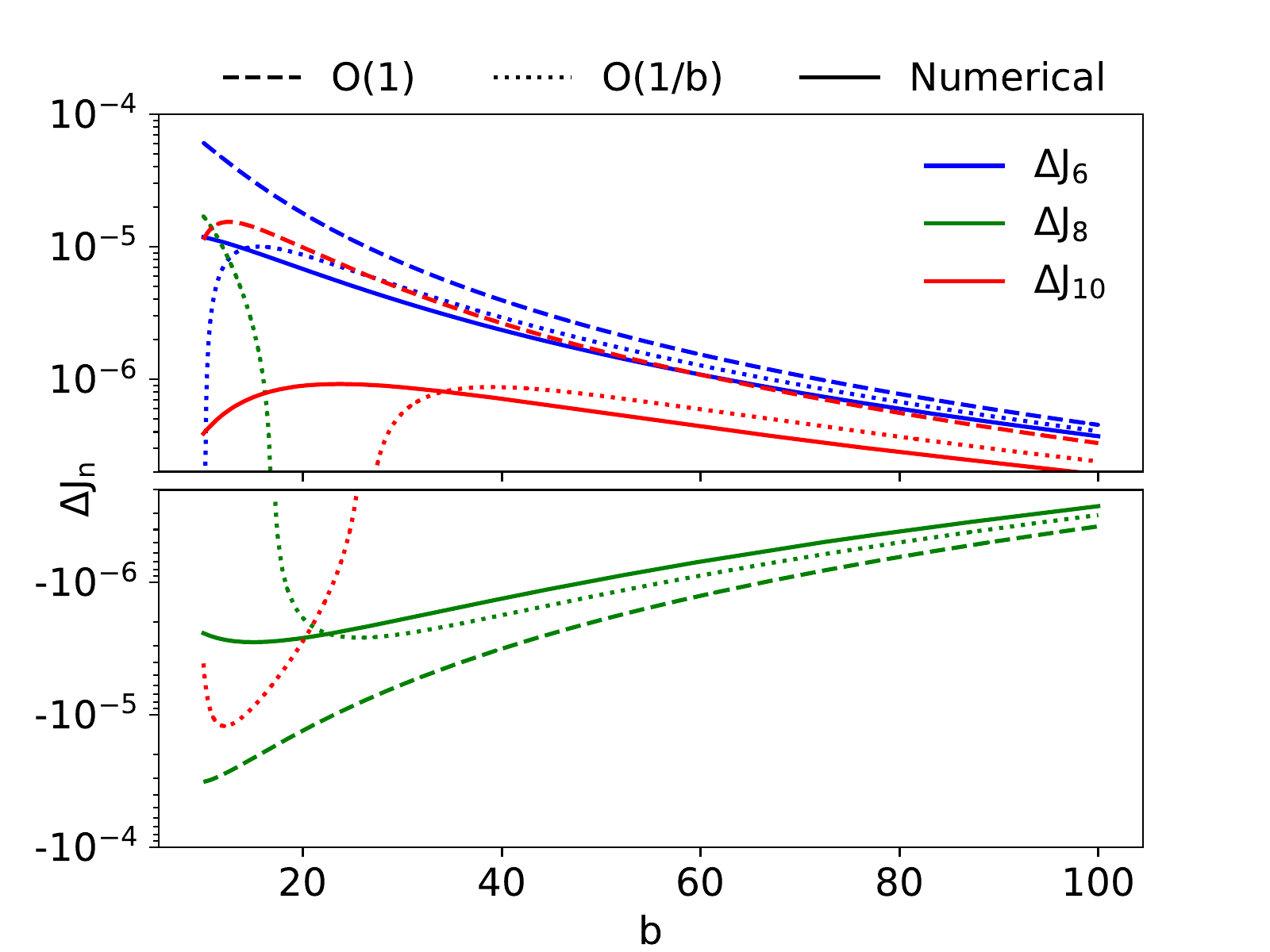}
\caption{A comparison of the results obtained from the analytical model with terms of the order O(1), (Equation~\ref{eq:O1order}), analytical model with terms of the order O(1/b) (Equation~\ref{eq:O1border}), and numerical calculations. The amplitude of differential rotation relative to solid body rotation, $a$, is chosen to be 0.04. The agreement between all the models only becomes good for large b. The value of $b$ at which agreement begins to improve is higher for higher order $n$ (as expected).  The analytical model is a rather poor approximation for small $b$ because we are calculating high order gravitational moments.}
\label{fig:deltajs}
\end{figure}

This result uses the saddle point method expansion (Appendix A) to first order in $O(1/b)$ (constant term + first order term) as the contributions from higher orders are negligible. If we consider only the lowest order term then we find that

\begin{equation}
\Delta J_{n} = C_n (-1)^{n/2+1} \; a \; q \; b^{-5/2}
\end{equation}

where $C_6 = 8.16$, $C_8 = 6.98$, and $C_{10} = 6.22$. However, these asymptotic (very small $d = 1/b$) limiting forms are not sufficiently accurate for the case of Saturn, as we shall discuss. The calculations from this analytical model are compared with the numerical calculations for a shallow flow that decays exponentially in Figure~\ref{fig:deltajs}. The agreement between the analytical model and the numerical calculation improves with the inclusion of the $O (1/b)$ term as well as for high $b$ (increasingly shallow flow). The flow depth at which agreement becomes good is also a function of the order of gravitational moment, as expected. This is simply because $\mathrm{cos} \theta$ changes a lot as $\mathrm{sin} \theta$ changes by a small amount, when $\theta$ is near $\pi/2$. Consequently, the Legendre polynomials of high order $n$ vary rapidly in the region of interest where the flow is non-negligible. The relationship between the flow properties and gravitational moments also becomes apparent in the plot. The deviations in gravitational moments decrease as the flow becomes shallower and as $n$ becomes larger.

\subsection{Effect of Planet's Oblateness} \label{sec:oblate}
We now address one of the key assumptions that we made in our model, i.e., the sphericity of the planet. Using our framework and perturbation theory, we can estimate the expected alteration in the calculated $\Delta J_n$ due to the non-spherical shape of the planet. To first approximation, the small deviation from sphericity can be captured with the second order Legendre polynomial:

\begin{equation}
r \rightarrow r \left(1 + \varepsilon P_2 \left( \mathrm{cos} \theta \right) \right)
\end{equation}

We Taylor-series expand the expression for gravitational moment $J_n$ (equation~\ref{eq:moment}), spherical Bessel function $j_n$ (equation~\ref{eq:homogeneous}), and $V_{pn}$ (equation~\ref{eq:V_p_projection2}) in the small quantity $\varepsilon$, which scales with $q$. The product of $P_2$ with $P_n$ can be written as a sum of three Legendre Polynomials, $P_{n-2}, P_{n},$ and $P_{n+2}$, where each of the polynomials is weighted by the corresponding Clebsch-Gordon coefficient \citep{Adams1878}. Expanding in small $\varepsilon$ and keeping the first order terms (see Appendix C), we get:

\begin{equation}
\Delta J_{n,oblate} \approx \Delta J_{n,sph}  - \varepsilon \bigg[ \frac{1}{4} (n+1) \Delta J_{n,sph} + \frac{3}{8} (n-1) \Delta J_{n-2,sph}  +  \frac{3}{8} (n+3) \Delta J_{n+2,sph}  \bigg] 
\end{equation}
where the subscript $sph$ stands for spherical i.e. values of $\Delta J_n$ calculated under the spherical planet approximation. The oblateness of a planet couples gravitational moments of different order.

Importantly, the effect of oblateness is small and of the order $\varepsilon \sim 0.1$ for Saturn. Another thing to note is that $\Delta J_{n-2}$ and $\Delta J_{n+2}$ are opposite in sign to $\Delta J_n$ for the flow profiles considered here\footnote{This does not necessarily have to be the case. The sign of $\Delta J_n$ depends on the functional form of differential flow $\epsilon$ and the decay depth adopted and can change when the depth is changed.} and the expression in the bracket is small. The effect of oblateness is roughly proportional to $n$, which implies that it has a greater effect on higher order $\Delta J_n$. This makes intuitive sense because higher order $\Delta J_n$ probe density in regions very close to the surface and therefore are more sensitive to the shape and oblateness of the planet.

Numerous studies have investigated the impact of assuming a spherical background state on $\Delta J_n$'s and compared their values from the (spherical) Thermal Wind equation and methods such as Concentric Maclaurin Spheroids (CMS) or Euler equation that account for the non-spherical shape of the planet \citep[e.g.][]{Kaspi2016, Cao2017a}. The differences in the $\Delta J_n$'s values from these two approaches do not exhibit the simple proportionality to the gravity moment degree $n$ that we propose above. We suspect this is because our perturbative approach breaks down for flows of moderate depth.

\section{Characterisation of Saturn's Differential Rotation}
\subsection{Numerical Calculations} \label{sec:calculation}

{\it Cassini's} measurements have revealed the deviations of higher order gravitational moments from solid body rotation and we adopt rough values of these in our work \citep{Iess2019}:

\begin{equation}
\Delta \mathrm{J}_6 \sim (5 \pm 0.5) \times 10^{-6} \; , \;  \Delta \mathrm{J}_8 \sim (-5.5 \pm 0.5) \times 10^{-6} \; , \; \Delta \mathrm{J}_{10} \sim (3.5 \pm 0.5) \times 10^{-6}
\end{equation}

What is remarkable about these values is their similarity in magnitude even though $n$ increases from 6 to 10. It is immediately clear that an exponentially decaying flow cannot match these observations for any reasonable value of $a$, the amplitude of differential flow (Figure~\ref{fig:deltajs}). We therefore must resort to another form for the flow profile.

The next natural step in formulating a flow profile is to consider the observed cloud-top motion as a surface manifestation of differential rotation in the planet's interior. There is no reason to suppose that winds and flows in the deep interior match the flow observed at the cloud-top level \cite[in fact, the observations indicate some vertical shear, see][]{Garcia-Melendo2011}. However, a sinc-like cylindrical flow profile that matches the surface flows of Saturn roughly reproduces the higher order gravitational moments surprisingly well \citep{Iess2019,Galanti2019} and also happens to have nice analytical properties (it is coincidentally the zeroth order Bessel function). In reality, it is not the lack of a differential flow profile that fits the data that has been the problem for Saturn. Degeneracy is the reason we cannot make conclusive statements about Saturn's flow properties using these measurements \citep{Iess2019,Galanti2019}.

Cloud-top motion extended deep into the planet as flows on cylinders \citep{Busse1976} is plausible provided the cylinders are external to the electrically conducting region. It is a useful working hypothesis due to the simplicity of the centrifugal potential that arises as a result of this flow. A more realistic model would include a radial depth dependent decay term \citep[e.g.][]{Galanti2017a,Galanti2019} but we cannot accommodate a radial dependence in our simple model because it violates Equation~\ref{eq:hydrostatic}. We therefore adopt a general sinc-like flow profile in our study.

For Saturn, differential rotation, as manifested on the surface and in the zonal flows, is known to be the order of a few percent of the solid body rotation rate ($\sim 4 \%$). This is a rough estimate and the exact value is not known. \footnote{This is due to the difficulty posed by the inference of the solid body rotation rate for Saturn \citep[see][]{Smith1982,Helled2015}, which still has some uncertainty and cannot be inferred from the highly axisymmetric magnetic field \citep{Cao2011}. This uncertainty is much smaller than the zonal flows however.} Given these uncertainties and the fact that flow in the interior may be different from cloud-top motion, we allow the amplitude and the depth of the flow to vary significantly from the values inferred from cloud-top motion.

\begin{figure}
\centering
\includegraphics[width=0.8\linewidth]{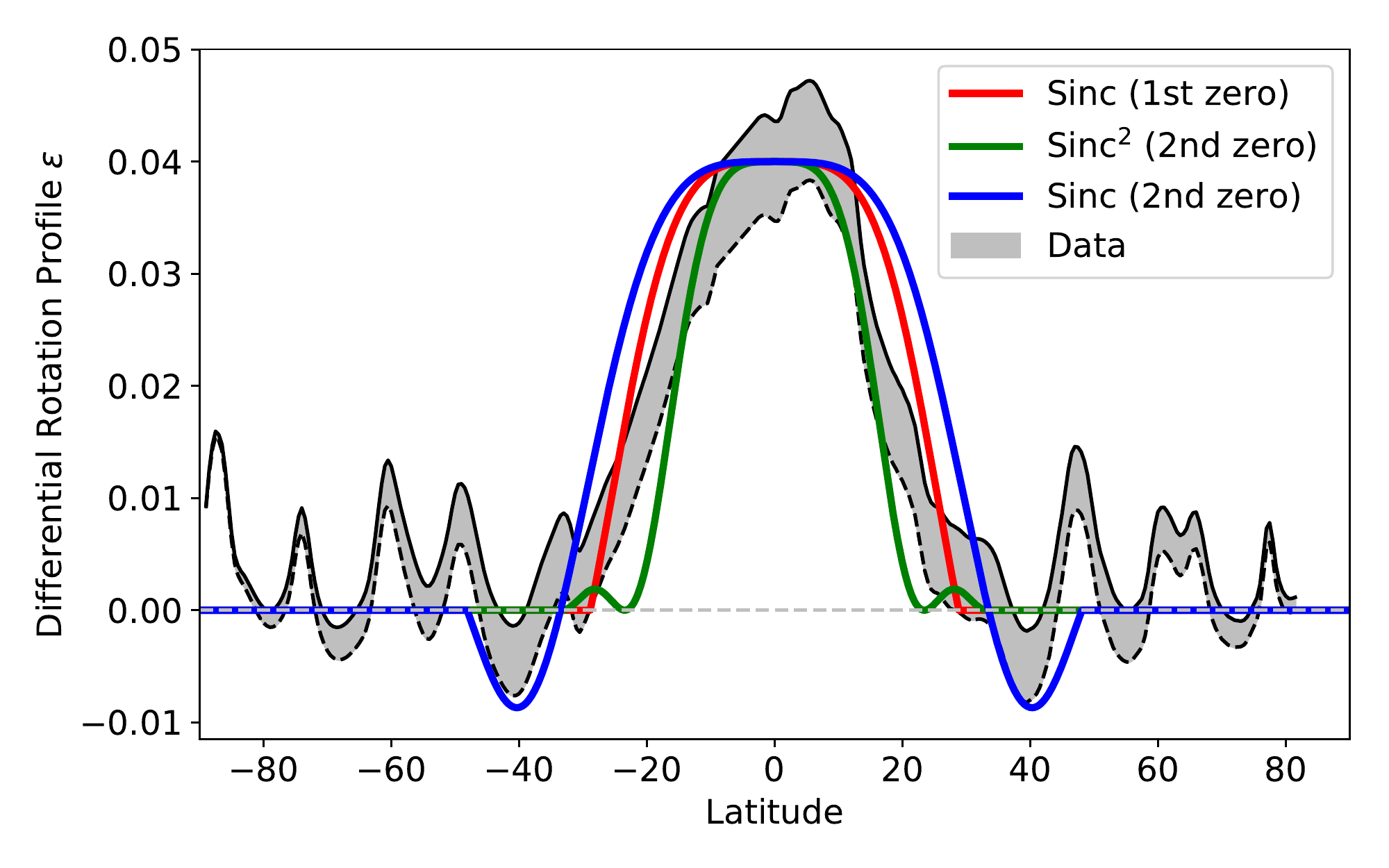}
\caption{Sample differential rotation profiles considered in this study: A sinc profile truncated at its first zero (red), a sinc profile with truncation at second zero (blue), and a sinc$^2$ profile truncated at second zero (green). The width and height of each oscillation is varied and corresponding deviations in gravitational moments are calculated. The cloud-top motions of Saturn are shown in black for two different rotation rates: 10h 39m 24s and 10h 34m 13s. \citep{Read2009}}
\label{fig:sample_profiles}
\end{figure}

Three broad functional forms based on the sinc function are chosen in the study (Figure~\ref{fig:sample_profiles}). The first is a sinc function truncated at its first zero with no flow interior to it.

\begin{equation}
\epsilon (s) = 
\begin{cases}
A \; \mathrm{sinc} \left( B \left(1-s \right) \right), & 1 -  \pi / B < s < 1 \\
0 , & 0 < s <  1 - \pi / B
\end{cases}
\label{eq:1stprofile}
\end{equation}

The second is a sinc squared functional form which only has prograde flow and is truncated at the position of second zero.

\begin{equation}
\epsilon (s) = 
\begin{cases}
A_1 \; \mathrm{sinc^2} \left( B_1 \left(1-s \right) \right), &  1 - \pi / B_1 < s < 1 \\
A_2 \; \mathrm{sinc^2} \left( B_2 \left(1-s \right) \right), &  1 - \pi / B_1 - \pi / B_2 < s < 1 - \pi / B_1  \\
0 , & 0 < s < 1 - \pi / B_1 - \pi / B_2 
\end{cases}
\label{eq:2ndprofile}
\end{equation}

The third form has the equatorial prograde flow and mid-latitude retrograde flow and the sinc function is truncated at the position of the second zero.

\begin{equation}
\epsilon (s) = 
\begin{cases}
A_p \; \mathrm{sinc} \left( B_p \left(1-s \right) \right), &  1 - \pi / B_p < s < 1 \\
A_r \; \mathrm{sinc} \left( B_r \left(1-s \right) \right) & 1 - \pi / B_p - \pi / B_r < s < 1-  \pi / B_p \\
0 , & 0 < s < 1 - \pi / B_p - \pi / B_r
\end{cases}
\label{eq:3rdprofile}
\end{equation}

For the second and third functional forms that have mid-latitude flow (prograde or retrograde), we allow the equatorial and mid-latitude flow to have independent amplitudes and depths. Although analytic forms are chosen for $\epsilon (s)$, subsequent calculations inadvertently involve numerical solutions and approximations. This is because analytical calculations for the sinc profile are time consuming. The entire calculation can be performed rapidly for exponential flow profiles. Therefore, we choose to represent the functional forms presented above as a linear combination of exponentials. A set of 15 exponentials (giving the desired accuracy) with a wide range of decay depths is chosen for this purpose:

\begin{equation}
\epsilon (s) = \sum_{i=1}^{15} \alpha_i e^{-2i (1-s)}
\label{eq:exponentials}
\end{equation}

The details of the numerical calculations are given in Appendix D. We now use this simple model to calculate gravitational moment deviations due to differential flow in Saturn and statistically characterize the dependence of these deviations on flow properties. This allows us to quantify the degeneracy in flow properties: a task that cannot be accomplished using the available suite of complex models due to their computational cost. Note that our approach does not work for Jupiter because it has significant hemispheric asymmetry, as indicated by the measurement of the odd moments \citep{Iess2018}.

\begin{figure}
\centering
\subfigure[]{\includegraphics[width=0.48\linewidth]{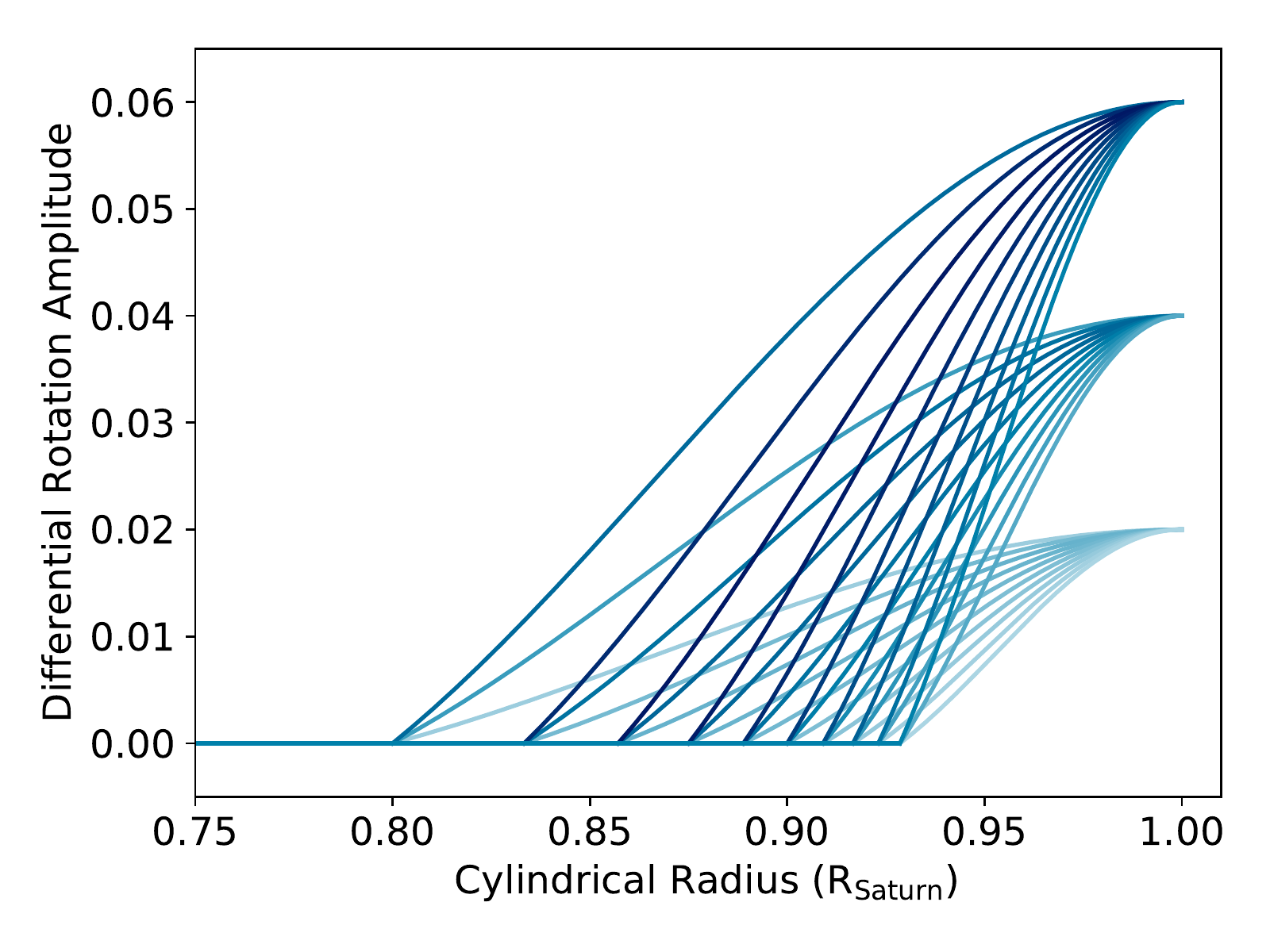}}
\subfigure[]{\includegraphics[width=0.48\linewidth]{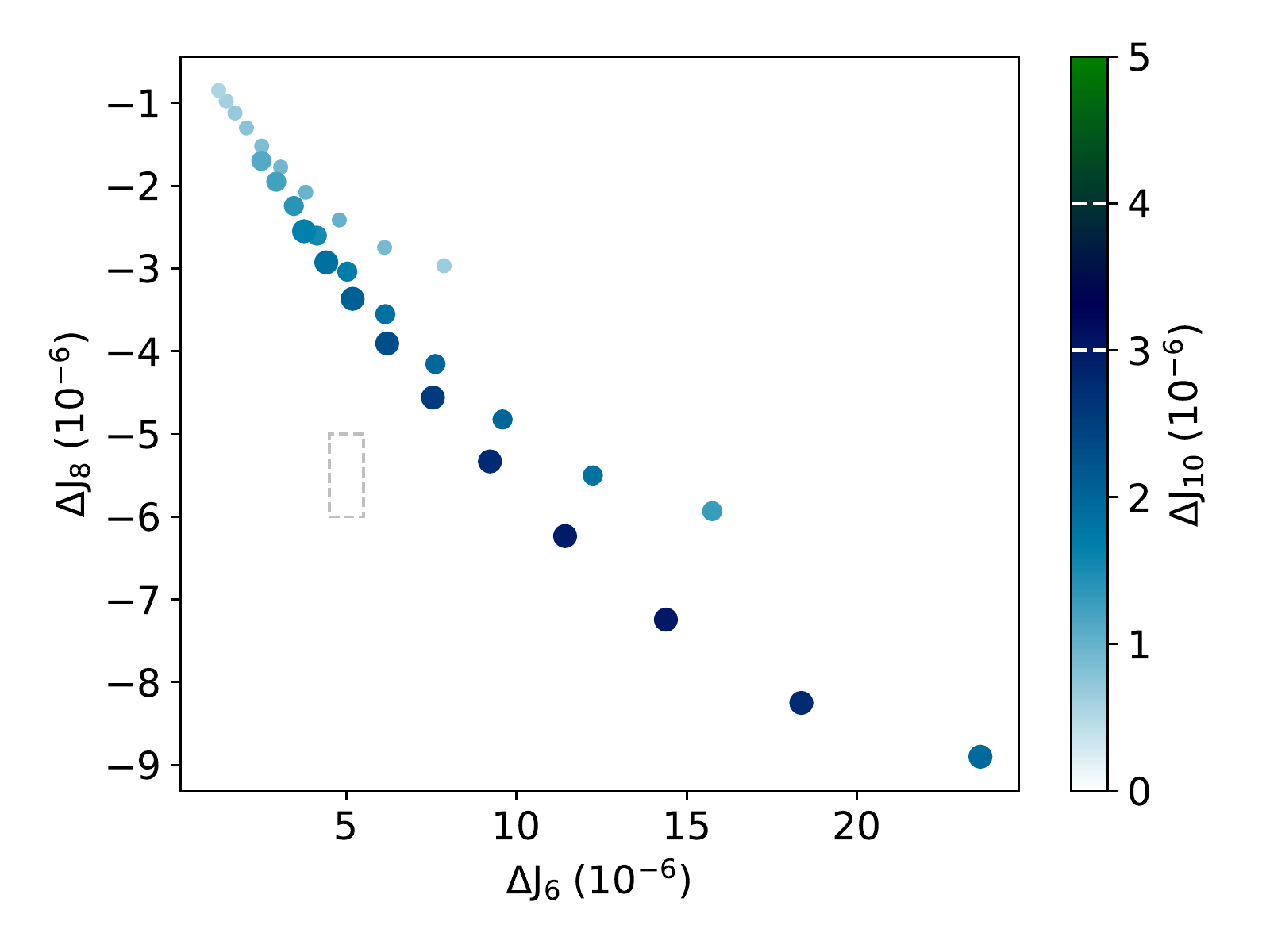}}
\caption{(a) Sinc profiles with varying amplitudes and depths truncated at the first zero that are used in this study (colored by $\Delta J_{10}$ values).
(b) The calculated values of $\Delta J_8$ plotted against $\Delta J_6$ and $\Delta J_{10}$ is indicated by color. The size of the scatter points corresponds to the three differential rotation amplitudes in (a). The grey dashed box indicates acceptable values of $\Delta J_n$ that agree with the observations. The magnitudes of $\Delta J_8$ and $\Delta J_{10}$ for a given $\Delta J_6$ are lower than they need to be to match the observations.}
\label{fig:sinc_1st_osc}
\end{figure}

\subsection{The Necessity of Mid-Latitude Retrograde Flow}

Efforts to fit the observed gravitational moments using highly accurate and sophisticated forward models have indicated that mid-latitude retrograde flow seems to be necessary to match the data \citep{Iess2019,Galanti2019}. We use our simple linear model to find out whether we reach the same conclusion or not. This serves as a test of both our model and the conclusion reached by other studies.

\begin{figure}
\centering
\subfigure[]{\includegraphics[width=0.48\linewidth]{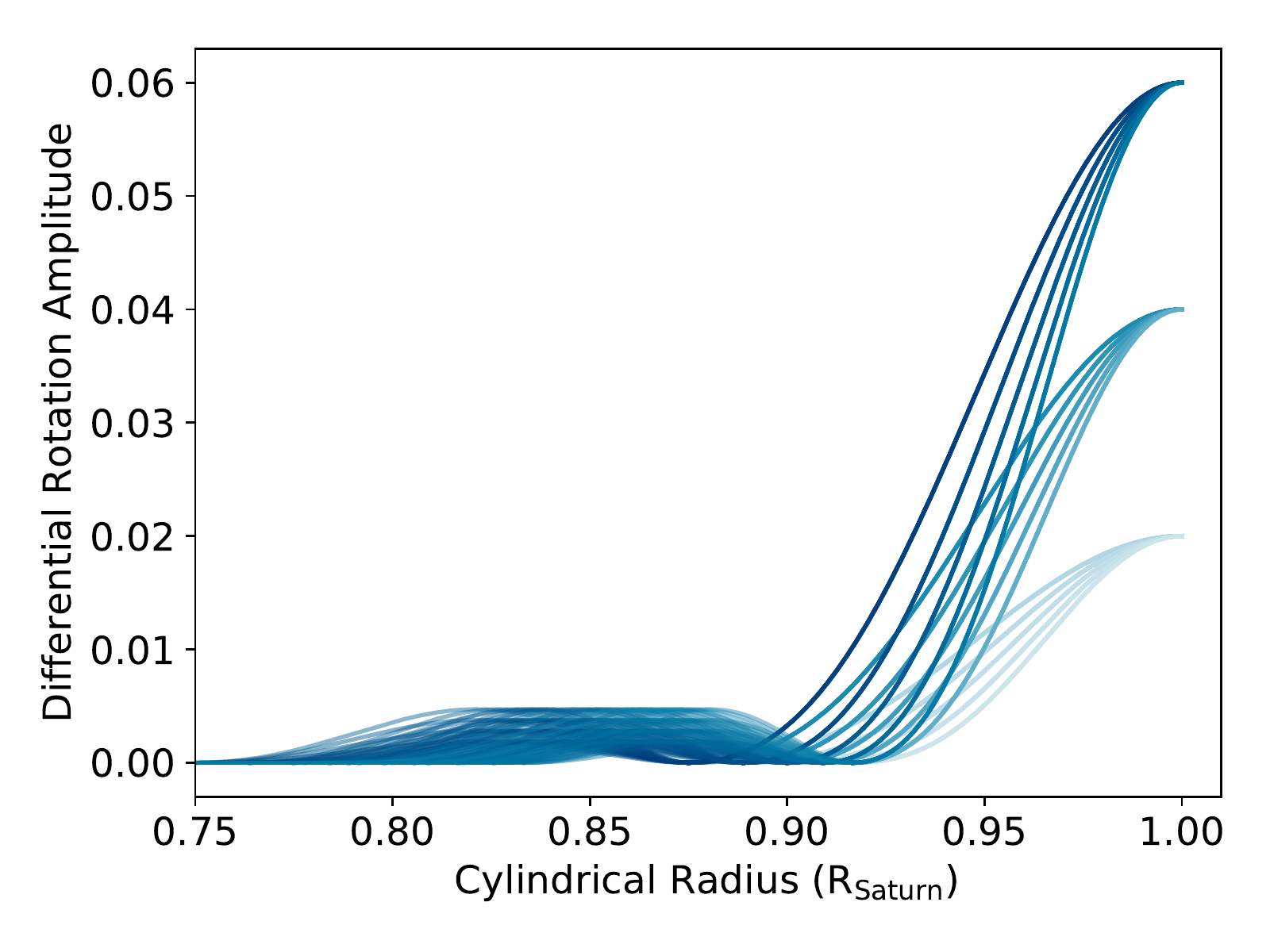}}
\subfigure[]{\includegraphics[width=0.48\linewidth]{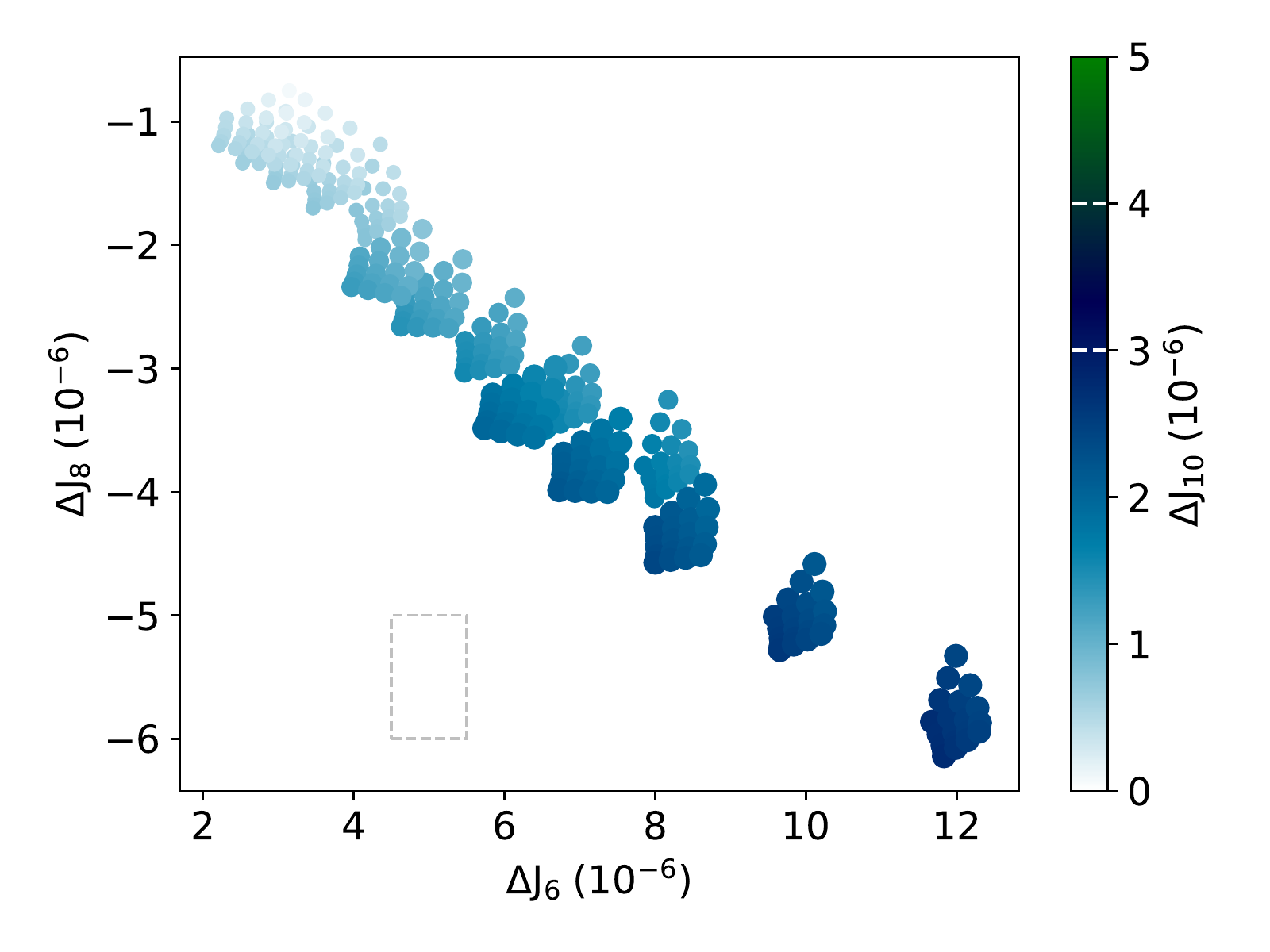}}
\caption{(a) Sinc squared profiles truncated at the second zero that are used in this study. The amplitude and depth of each oscillation are allowed to vary to determine how the $\Delta J_n$ values depend on them.
(b) The calculated values of $\Delta J_8$ plotted against $\Delta J_6$ and $\Delta J_{10}$ is indicated by colour. The size of the scatter points corresponds to the three differential rotation amplitudes in (a). The grey dashed box indicates acceptable values of $\Delta J_n$ that agree with the observations. It is clear that none of the amplitudes or depths match the observations with such a differential flow profile.}
\label{fig:sinc_2_osc}
\end{figure}

The deviations in gravitational moments due to prograde flows (equatorial and/or mid-latitude; defined in Equation~\ref{eq:1stprofile} and \ref{eq:2ndprofile}) are shown in Figure~\ref{fig:sinc_1st_osc} and \ref{fig:sinc_2_osc}. The different flow profiles are plotted in the left panels of the figures and the corresponding values of $\Delta J_n$ are shown in the right panels. The dashed lines in the plot mark the measured values of the gravitational moments. None of the flow profiles agree with the observations. The calculated $\Delta J_n$ for prograde flows exhibit behaviour that is similar to those obtained for exponentially decaying profiles. That is, the prograde flow are unable to produce $\Delta J_n$ of similar magnitude for $n = 6, 8,$ and 10. We therefore conclude that for cylindrical differential flow profiles, the absence of a retrograde flow at intermediate latitudes makes it difficult to fit the observed values of $\Delta J_n$. This is in agreement with other studies that use more sophisticated models.

\subsection{Non-Uniqueness of Flow Properties}
Given that a prograde flow is not able to reproduce the observed $\Delta J_n$ values, we focus on our third general flow profile: a sinc function truncated at its second zero (Equation~\ref{eq:3rdprofile}). This function is characterised by four parameters: the prograde flow depth and amplitude ($D_p$ and $A_p$) and the retrograde flow depth and amplitude ($D_r$ and $A_r$)\footnote{The depth and amplitude are normalised by Saturn's radius and solid body rotation amplitude respectively}. To understand the degeneracy in flow properties, we use affine-invariant Markov Chain Monte Carlo (MCMC) simulations to constrain the parameters that characterize Saturn's differential flow. This is made possible due to our choice to write all flow profiles as a linear combination of the same set of exponentials (Equation~\ref{eq:exponentials}). The gravitational moments due to each of these exponentially decaying flows is calculated in advance and for each flow profile, we simply sum the weighted contribution of these moments (see Appendix D). The widely used Python package {\it emcee} is used to perform the MCMC simulations \citep{Foreman-Mackey2013}. We assume uniform priors for the four model parameters that characterize the data flow and plot the corresponding posterior probability distributions from the MCMC analysis in order to determine the information provided by these new data.

As mentioned before, the chosen range of parameters is inspired by the observed surface zonal wind profile. However, we allow for variation in flow properties to account for the possibility that surface flows do not trace the flows in the interior. We use 20 `walkers' and 35,000 steps in each chain, which gives us a total of 700,000 samples of the posterior probability distribution. We initialize our walkers at randomly chosen positions clustered near a solution that is in reasonably good agreement with the data. We then run the MCMC for an initial 1000 step burn-in phase in order to ensure that all walkers have reached the preferred region of parameter space. The integrated auto-correlation time ($\tau_{f}$) for each of the model parameters is $\sim 100 - 150$. This implies that we draw $700,000/\tau_{f} \sim 4500$ independent samples from the posterior distribution, indicating that we have adequately sampled the relevant parameter space \citep{Foreman-Mackey2013}.

\begin{figure}
\centering
\includegraphics[width=0.8\linewidth]{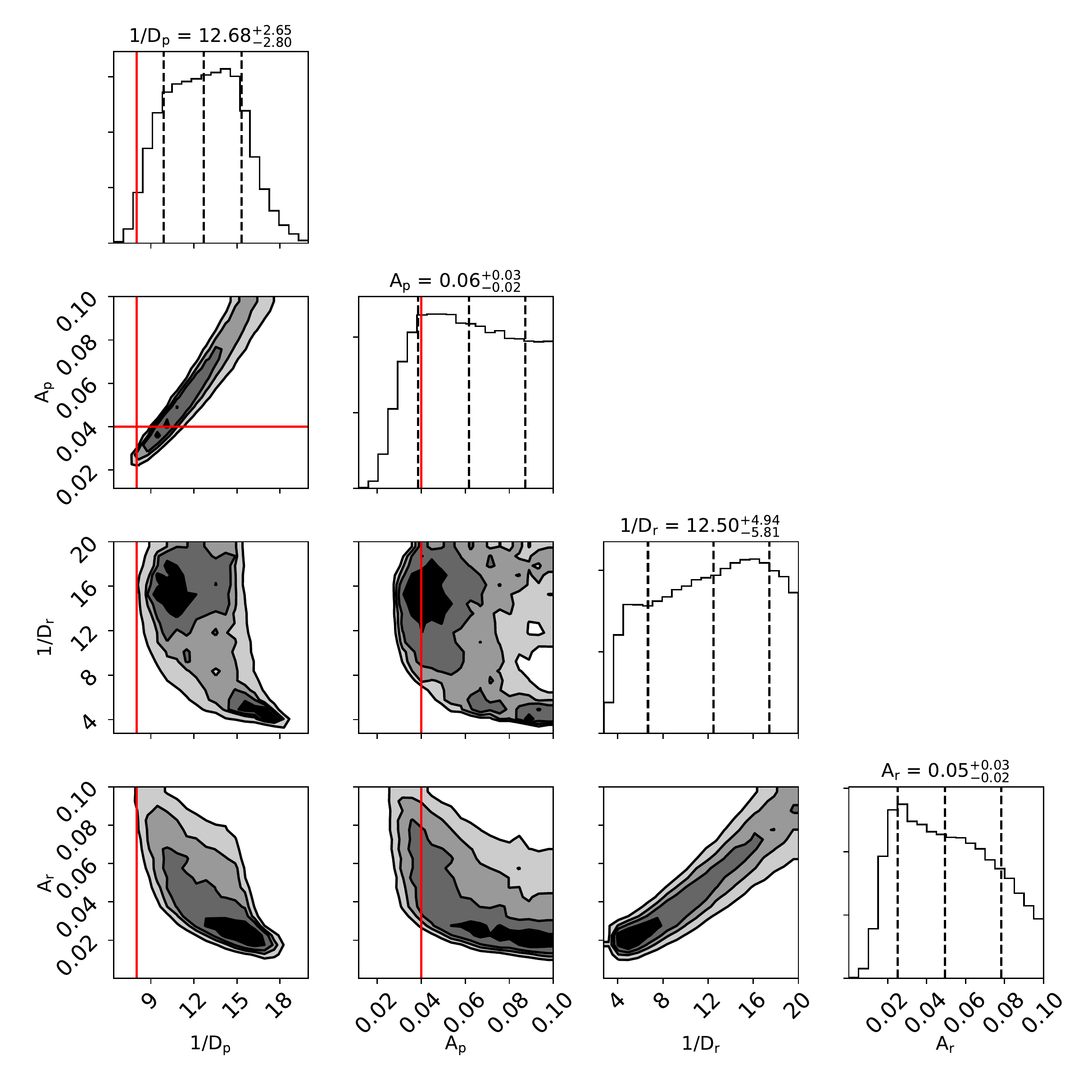}
\caption{Posterior probability distribution for the four parameters used in our model. The red line indicates parameters roughly corresponding to the equatorial prograde cloud-top motion of Saturn, i.e. a $\sim$ 4 \% flow amplitude with respect to the background rotation rate and a latitudinal extent of $\pm 30^{\circ}$ \citep{Read2009}. The black dashed lines mark the 16th, 50th, and 84th quantiles for the 1D histograms. We note that there is a large degeneracy between the flow depth and amplitude for both the prograde and the retrograde flow. In addition, there is some anti-correlation between the properties of prograde and the retrograde flow. The posteriors for each parameter show that a wide range of values yield acceptable gravitational moments given the data.}
\label{fig:sinc_4param_MCMC}
\end{figure}

The posterior probability distributions for the prograde and retrograde flow amplitude and depth are shown in Figure~\ref{fig:sinc_4param_MCMC}. This corner plot gives us the posteriors for each of these parameters as well as the correlations for all pairs of parameters. The surface (cloud-top) flow amplitude and the extent of the prograde jet on Saturn are marked by red lines in these plots. One thing that becomes immediately evident is that the data do not constrain the parameters very tightly. The posteriors for the parameters taper off at the edges of the parameter space but are not that different from the assumed flat prior for the MCMC simulations. This implies that a wide range of flow parameters can reasonably explain the observed gravitational moments, given the errors in our measurements.

Correlations between flow parameters allow us to understand the non-uniqueness of the possible solutions and place greater constraints than the 1D histograms. We find that the flow depth and amplitude for both the prograde and retrograde flow are strongly correlated. A shallow flow with a large amplitude and a deep flow with a small amplitude produce similar values for the deviations in gravitational moments. Moreover, we note that correlations between properties of the prograde and retrograde flow are weak. The amplitude of the prograde flow and the depth of the retrograde flow are almost uncorrelated. As for the other parameter pairs, we find that the depth and the amplitude of prograde and retrograde flow are {\it roughly} anti-correlated with each other (the actual posterior distribution is not simple). That is, the retrograde flow tends to be shallow if the prograde flow is deep (and vice versa) and the amplitude of the retrograde flow tends to be smaller if the prograde flow is strong (and vice versa).

\begin{figure}
\centering
\subfigure[]{\includegraphics[width=0.55\linewidth]{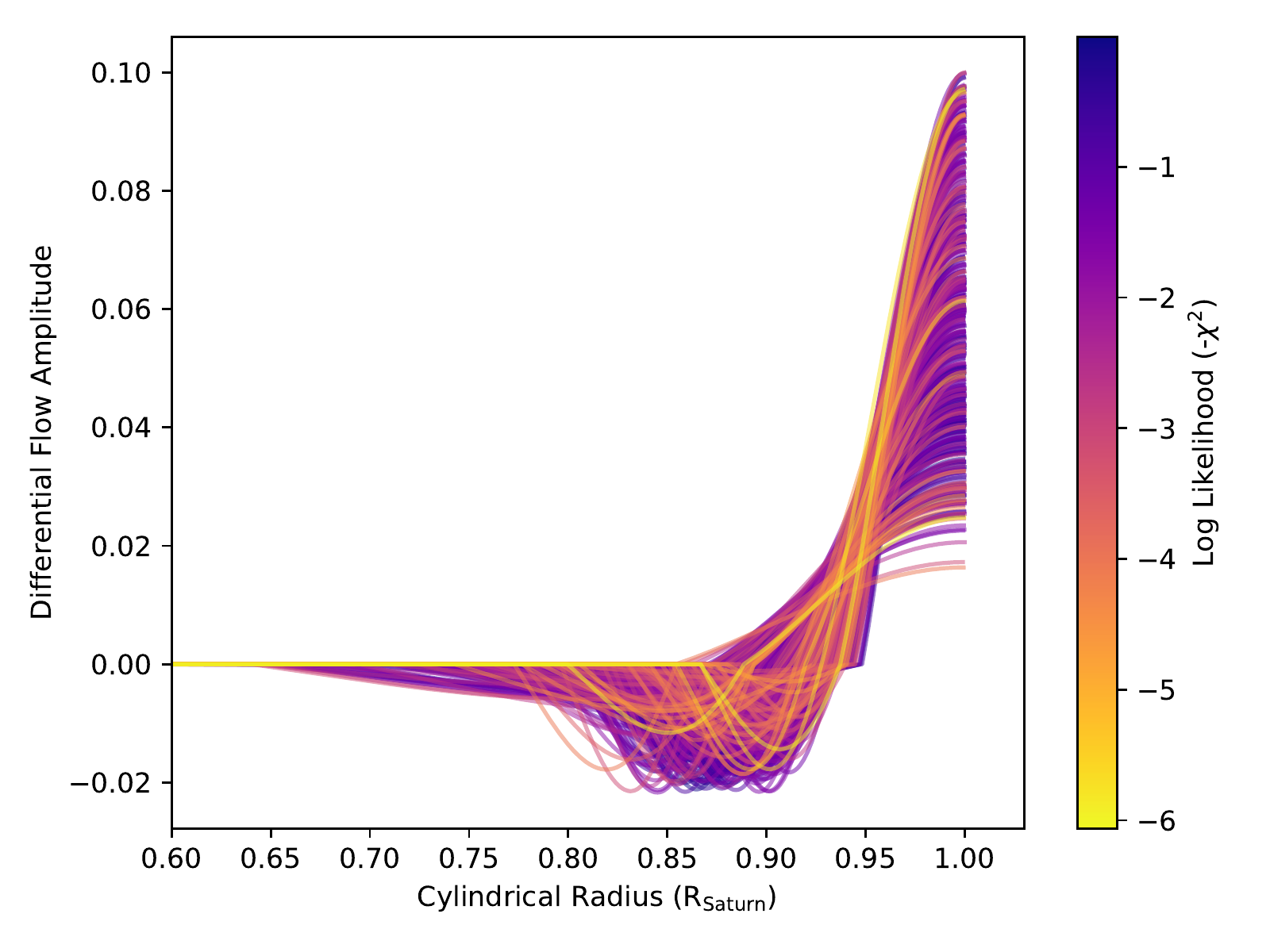}}
\subfigure[]{\includegraphics[width=0.49\linewidth]{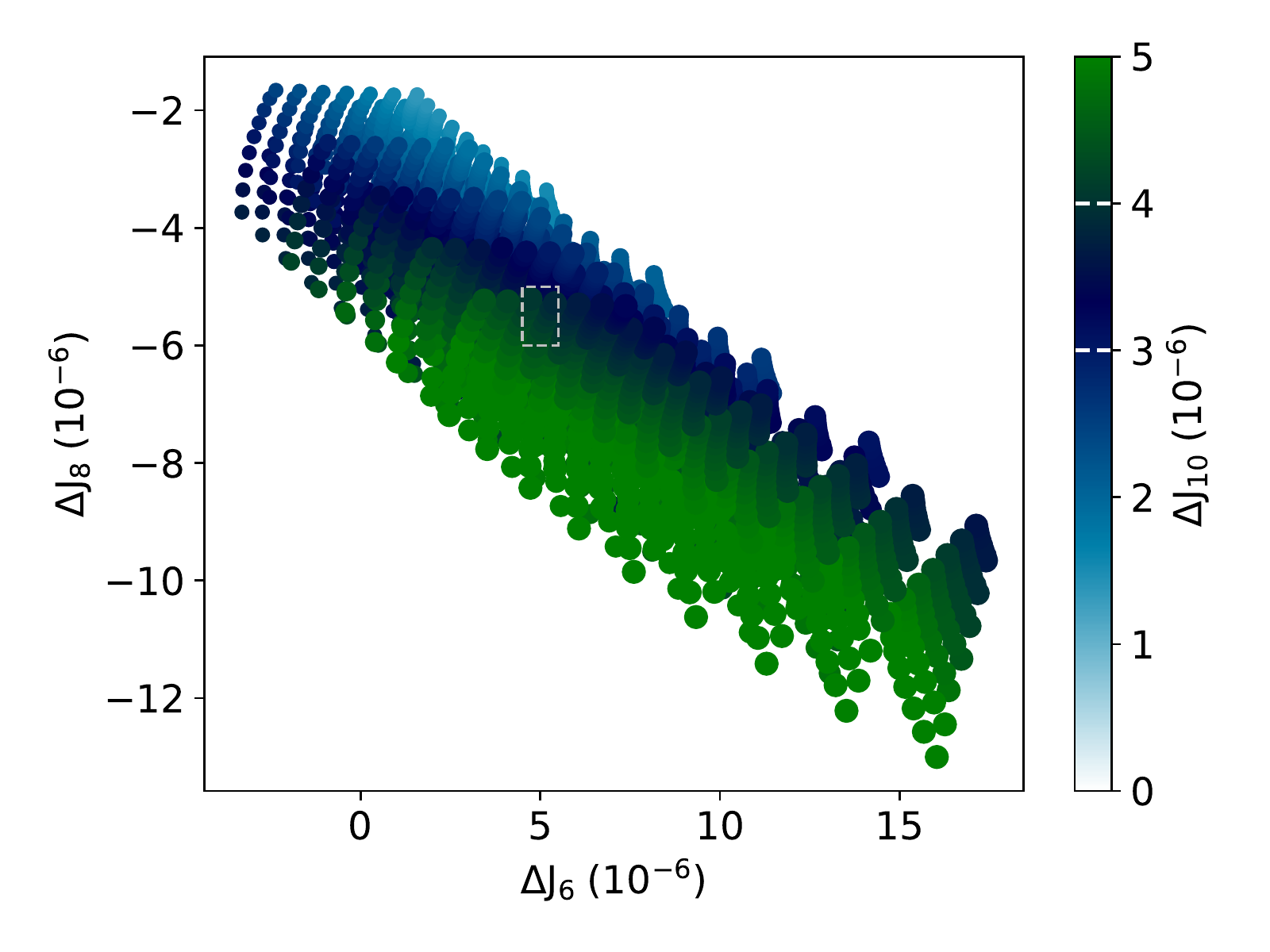}}
\subfigure[]{\includegraphics[width=0.49\linewidth]{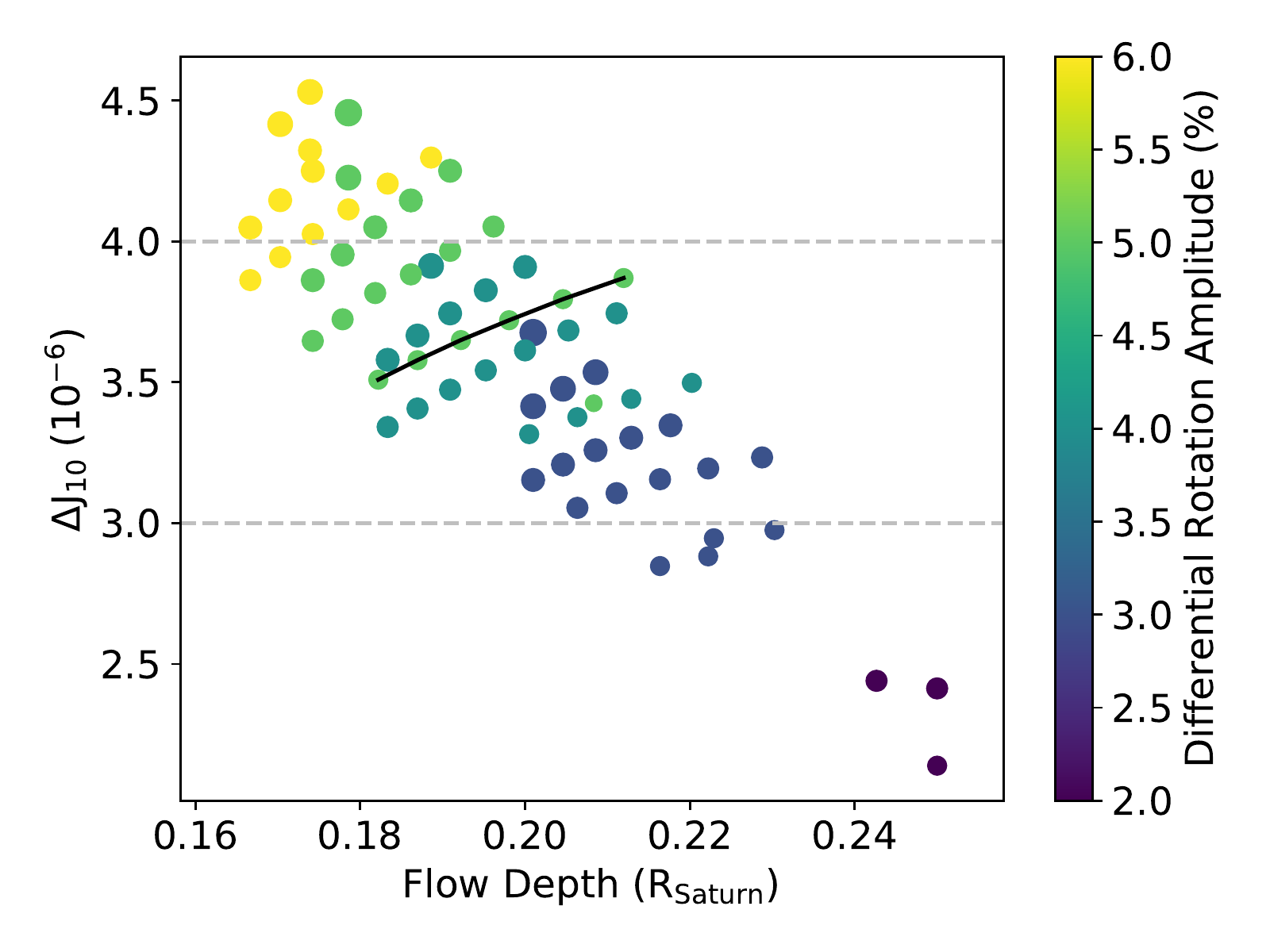}}
\caption{(a) A sample of profiles corresponding to MCMC runs colored according to their log likelihood values. The profiles with the largest log-likelihood (smallest $\chi^2$) best fit the data.
(b) The inclusion of retrograde flow at mid-latitudes enables us to match the data and model. The region of interest is densely populated with models with varying differential flow properties.
(c) All the plotted points here correspond to model properties that agree with the observed values of $\Delta J_8$ and $\Delta J_6$. The grey dashed lines demarcate acceptable values for $\Delta J_{10}$. $\Delta J_{10}$ is plotted against the flow depth ($D_p + D_r$) and the color indicates the flow amplitude of the prograde flow. The size of the scatter plots is a proxy for the amplitude of retrograde flow. The black line traces models that have all properties constant apart from the depth of the retrograde flow. A deeper and stronger retrograde flow yields a higher $\Delta J_{10}$ value.}
\label{fig:sinc_4param}
\end{figure}

We plot a sample of flow profiles, colored according to their log-likelihood ($- \chi^2$), for which the MCMC calculations are performed in Figure~\ref{fig:sinc_4param}a. These profiles are a part of the MCMC `chain' as the simulation explores the parameter space to fit the observed gravitational moments. The profiles with the highest log-likelihood (equivalent to the smallest $\chi^2$) best fit the observations. Smaller amplitude flows that are deep tend to fit the observations as well as larger amplitude shallow flows.

Inclusion of retrograde flow clearly improves the agreement between the data and the model (Figure~\ref{fig:sinc_4param}b). It is therefore worth noting how the properties of retrograde flow influence the gravitational moments. This will help us understand why including retrograde flow seems necessary to fit the data. Essentially, models with only prograde flow that match $\Delta J_6$ and $\Delta J_8$ underestimate the value of $\Delta J_{10}$ (see Figure~\ref{fig:sinc_1st_osc} and \ref{fig:sinc_2_osc}). Including retrograde flow and increasing its amplitude or depth increases $\Delta J_{10}$ while leaving the other two gravitational moments relatively unaltered (Figure~\ref{fig:sinc_4param}c). This is what allows us to obtain a reasonable match between the model and the observations.

\section{Discussion and Conclusions}
The ideas and theory presented here aid in the development of an intuitive understanding of how the properties of the differential rotation are related to the deviations (from solid body rotation) in gravitational moments. Gravitational moments are given by the integral of density and a Legendre polynomial of certain order. In our formalism, density and gravitational potential are related simply: $2 K \rho = V + V_{rot}$. The contribution due to differential rotation, $\delta V_{rot}$, therefore directly relates to the density, and hence to the gravitational moments, which are now given by the integral of gravitational potential with the Legendre polynomial. Additionally, $\delta V_{rot} = q \int_{0}^{s} 2 \epsilon(s') s' \mathrm{d}s'$ is related to cylindrical flow properties contained in $\epsilon$. This establishes a somewhat more comprehensible relation between differential rotation properties and the gravitational moments.

Using this model, we have shown that deviations in gravitational moments alone do not suffice to constrain the differential flow properties of Saturn very accurately due to the errors in the measurement and the modeling. Even using a single sinc-like cylindrical flow profile, which already severely restricts the parameter space, we are able to match the observed gravitational moments for a wide range of flow properties. There is added non-uniqueness in the functional form of the flow profile as well and the surface flows, although coupled to the interior dynamics, may be susbtantially different from those in the interior \citep[e.g.][]{Kong2018}. To deal with this non-uniqueness, we must seek some other ways of constraining the flow depth or amplitude which will narrow down the range of possible flow properties. This requires invoking additional physics that might place constraints on the differential flow of Saturn.

One promising way of constraining flow depth is by studying the interaction of interior differential flow and planetary magnetic fields. Electric conductivity rises steadily as one ventures deep into the planet's interior and any differential flow would lead to the generation of currents and Ohmic dissipation. In the deep interior, the conductivity is very high and any differential rotation is quickly damped out by MHD drag. However, in the semi-conducting region, it is possible to have small amplitude flows with speeds of $\sim$ 1 cm/s to 1 m/s. Electrical conductivity and Ohmic dissipation along with the measured interior heat flux of the planet place this upper limit on the amplitude of differential flow because strong flows would generate a large amount of heat, which would lead to a discrepancy between the expected and the measured heat flux. \cite{Cao2017b} studied this interaction in the semi-conducting region of Jupiter and Saturn to determine if flow induced magnetic field variations can be measured with the Juno mission or the {\it Cassini} grand finale orbits. Their work indicates that for an intermediate electrical conductivity $\sigma \approx 10^{-2}$ S/m, it is not possible to have large amplitude (100 m/s) flows. However, 1 m/s flows are possible. The electrical conductivity reaches this value at $\sim$ 0.85 R$_{\mathrm{Saturn}}$, which implies that the differential flow must decay from 100s of m/s at the surface to a negligible 1 m/s at this radius. This could be used as a constraint on the flow depth in our efforts to determine the differential flow properties. Indeed, the measured gravity moments of Saturn seem to imply an upper limit on the flow depth that is consistent with this constraint placed by MHD \citep{Iess2019, Galanti2019}. Notably, Ohmic dissipation places a constraint on the {\it radial} depth of the flow, not just the cylindrical depth $r_c$ considered in this work. It is therefore possible to have differential rotation at high latitudes as long as it becomes negligible at this transition radius.

Using such a constraint and our results (in particular, Figure~\ref{fig:sinc_4param_MCMC} and \ref{fig:sinc_4param}), we note that deep flows with small amplitudes are excluded and shallow flows with large amplitudes are preferable because they agree with the gravity data as well as the theoretical expectations from magnetohydrodynamics and Ohmic dissipation. More specifically, the flow depth needs to meet the following conditions: $1/D_p + 1/D_r \lessapprox 25$ with both $ 1/D_p \geq 6$ and $1/D_r \geq 6$. The required amplitude of the flow would depend on the flow depths chosen and it can vary from $\sim 4 - 8\%$.

In conclusion, we have presented a linear model that relates differential flow to gravitational moments in this paper. We demonstrated its utility in the development of an intuitive understanding of the underlying phenomenon and the rapidity of the calculations that enable a statistical analysis of the flow parameters. Our analytical calculations for a shallow flow are useful for understanding the relationship between flow properties and resulting gravitational moments. In addition, we have derived an expression that quantifies the effect of a planet's oblateness on its gravitational moments, providing first order corrections to the assumption of sphericity that we made to make our model linear. This expression will prove useful in comparing the effect of oblateness with other small corrections to the gravitational moments. Moreover, the linearity of the model allows us to make a quick forward calculation for the gravitational moments given a (cylindrical) flow profile. We used this feature of our model to constrain the properties of Saturn's differential flow based on the higher order gravitational moments measured by {\it Cassini} by performing MCMC simulations. This calculation allowed us to quantify the widely suspected non-uniqueness of Saturn's flow parameters inferred from the gravity data. We found that retrograde flow seems necessary to explain the observed gravitational moments as flows that are purely prograde tend to underestimate $\Delta J_{10}$. A wide range of flow parameters yield gravitational moments that agree with the observations and the flow depth and amplitude are anti-correlated. Given the non-uniqueness of the flow properties, additional physics needs to be invoked to place tighter constraints on them. Matching the gravitational moments along with the theoretical expectations from Ohmic dissipation due to flow in the semi-conducting region of Saturn and its heat flux might provide an additional way of diminishing the allowed parameter space of differential flow properties.

\section*{Acknowledgements}
Y.C. thanks Hao Cao for stimulating discussions, Steve Markham for help with {\it Mathematica} \textcopyright, and Heather Knutson for discussions on MCMC. We are grateful to the reviewers for their thoughtful comments and suggestions which improved this manuscript. This work uses matplotlib \citep{Hunter2007}, scipy \citep{scipy}, and sympy \citep{Meurer2017}.

\bibliographystyle{apalike}
\bibliography{main}

\appendix
\section{Saddle Point Method}
The saddle point method is used for estimation of integrals of functions with very sharp peaks and it was first published by \cite{Debye1909}. For an integral of the form below for real functions $f(x)$ and $g(x)$ and for large $A > 0$

\begin{equation}
I (A) = \int_{x1}^{x2} f(x) e^{Ag(x)} \mathrm{d}x
\end{equation}

Since the function is strongly peaked at some $x = x_\mathrm{o}$ for large $A$, we can approximate the integral as:

\begin{equation}
I (A) \approx f(x_\mathrm{o}) e^{Ag(x_\mathrm{o})} \sqrt{\frac{2 \pi}{-A g''(x_\mathrm{o})}} \bigg( 1 + \sum_{n=1}^{\infty} \frac{C_{2n}}{A^n} \bigg)
\end{equation}

The C coefficients depend on the derivatives of $f (x)$ and $g (x)$. We include $C_2$ in our model and it is given by: $C_2 = -f''/4fg'' + f' g''' / 8 f (g'')^2 + g'''' / 32 (g'')^2 - 5 (g''')^2 / 192 (g'')^3$.

\section{Shallow Flow: Detailed Calculation}
Let us assume that the differential flow profile can be represented by a steeply declining exponential which is only a function of the cylindrical radius of a fluid planet:

\begin{equation}
\epsilon (s) = a \; e^{-b \; (1-s)}
\end{equation}

Here, $a$ and $b$ are constants and $s$ is the cylindrical radius (as a fraction of the equatorial radius). We can integrate this function analytically to obtain $\delta V_{rot}$:

\begin{equation}
\delta V_{rot} (s) = 2 q \int_{0}^{s} \epsilon (s') \; s' \; \mathrm{d}s' = a \; q \; e^{-b} \; \frac{ 1 + e^{bs} (b s - 1)}{b^2}
\end{equation}

Solving the equation for $V_p$ (Equation 6) and using the condition that $V_p$ and its derivative are zero at $s = 0$ to set the constants:

\begin{equation}
V_p (s) = -  \frac {2 a  \; q \; e^{-b} k^2} {b^3} \bigg[   \frac{b s^2}{2}  + e^{b s} \bigg( s - \frac{3}{b} \bigg) + 2s + \frac{3}{b}  \bigg]
\end{equation}

To proceed further with the analytic calculation, we need to project $V_p$ onto spherical coordinates using the Legendre Polynomials (Equation 7 and 8). For very shallow flow, we can use a Taylor series expansion of the Legendre Polynomials up to a certain order (that we choose) and calculate the integrals analytically. We choose to expand the even Legendre Polynomials to second order in $\mathrm{cos} \; \theta$ around $\theta = \pi/2$:

\begin{equation}
P_n (\mathrm{cos} \; \theta) = {n \choose n/2} \frac{(-1)^{n/2} }{2^n} \bigg[ 1 - \frac{n \; (n+1)}{2} (\mathrm{cos} \theta)^2 \bigg]
\end{equation}

With this approximation, we can project $V_p$ onto a basis set of Legendre Polynomials with $s = r \; \mathrm{sin} \theta$. We then have an expression for $V_{pn}$ from Equation 8:

\begin{multline}
V_{pn} (r) = - \frac{2n + 1}{2} {n \choose n/2} \frac {2 a  \; q \; e^{-b} k^2} {b^3} \frac{(-1)^{n/2} }{2^n} \;  \times \\ 
\int_{-1}^{1} \bigg[   \frac{b (r \; \mathrm{sin} \theta)^2}{2}  + e^{b r \; \mathrm{sin} \theta} \bigg( r \; \mathrm{sin} \theta - \frac{3}{b} \bigg)  + \frac{3}{b} + 2 r \; \mathrm{sin} \theta \bigg] \times  \\
\bigg[ 1 - \frac{n \; (n+1)}{2} (\mathrm{cos} \theta)^2 \bigg] \mathrm{dcos}\theta
\end{multline}

We calculate this integral in the approximation of shallow flow such the exponential approaches a delta function. In this limit, we can neglect all terms except the exponential in the expression for $V_p$ and the expression for $V_{pn}$ simplifies to:

\begin{multline}
V_{pn} (r) \approx - \frac{2n + 1}{2} {n \choose n/2} \frac {2 a  \; q \; e^{-b} k^2} {b^3} \frac{(-1)^{n/2} }{2^n} \;  \times \\ 
\int_{-1}^{1} e^{b r \; \mathrm{sin} \theta} \bigg( r \; \mathrm{sin} \theta - \frac{3}{b} \bigg)
\bigg[ 1 - \frac{n \; (n+1)}{2} (\mathrm{cos} \theta)^2 \bigg] \mathrm{dcos}\theta
\end{multline}

To evaluate this integral, we resort to the saddle point method, which works well for strongly peaked functions such as the exponential in the expression above.

\begin{multline}
V_{pn} (r) \approx - \frac{2n + 1}{2} {n \choose n/2} \frac {2 a  \; q \; e^{-b} k^2} {b^3} \frac{(-1)^{n/2} }{2^n} \;  \bigg( r - \frac{3}{b} \bigg) \sqrt{\frac{2 \pi}{b r}} e^{b r} \times \\
\bigg[1  - \frac{1}{4br} \bigg( \frac{1}{r-\frac{3}{b}} + n(n+1)+\frac{3}{8} \bigg) + O  (1/b^2r^2) \bigg] 
\end{multline}

Here, the first term in the parenthesis is the O(1) term which does not depend on $b$ or $r$. The second term is the order O(1/b) expansion in the saddle point approximation. We calculate the expression for $\Delta J_n$ for both these orders of expansion below. Evaluating $V_{pn}$ and its derivative at $r = R = 1$ (fractional radius in our calculations) gives us our final expression for $\Delta J_n$:

\begin{multline}
\Delta J_n = - \frac{2n + 1}{2} {n \choose n/2} \frac {2 a  \; \Omega_{\mathrm{o}}^2 \; \pi^2} {b^3} \frac{(-1)^{n/2} }{2^n} \sqrt{\frac{2 \pi}{b}}  \;
 \frac{\bigg[ -\frac{5}{2} + b + \frac{3}{2b} \bigg] j_n(\pi) - \pi \bigg[ 1 - \frac{3}{b} \bigg] j'_n(\pi)}{(n+1)j_n(\pi) + \pi j'_n(\pi)}
\label{eq:O1order}
\end{multline}

\begin{multline}
\Delta J_n = - \frac{2n + 1}{2} {n \choose n/2} \frac {2 a  \; q \; \pi^2} {b^3} \frac{(-1)^{n/2} }{2^n} \sqrt{\frac{2 \pi}{b}}  \; \times \\
\frac{\bigg[ \left( -\frac{5}{2} + b + \frac{3}{2b} \right) + \frac{n (n+1)}{4b} \left( \frac{7}{2} -b - \frac{9}{2b} \right) \bigg] j_n(\pi) - \pi \bigg[ \left(1 - \frac{3}{b} \right) \left( 1 - \frac{n(n+1)}{4b} \right) \bigg] j'_n(\pi)}{(n+1)j_n(\pi) + \pi j'_n(\pi)}
\end{multline}

\section{Effect of Planet's Oblateness}
We discussed the consequence of relaxing our assumption about the spherical shape of the planet on the gravitational moments of a celestial body. Here, we show the calculation in detail instead of simply quoting the results. The result we present is essentially derived from the continuity of the gravitational potential at the surface of the planet, which is no longer spherical but is as described by:

\begin{equation}
r \rightarrow r \left(1 + \varepsilon P_2 \left( \mathrm{cos} \theta \right) \right)
\end{equation}

Here, $\varepsilon$ is small (typically $<< 1$)and $P_2$ is the second order Legendre polynomial. The boundary condition for the potential at the surface is:

\begin{equation}
\sum_{Even \; n} V_{pn} (r) P_n (cos \theta) + \sum_{Even \; n} A_n j_n (kr) P_n (cos \theta) = - \sum_{Even \; n} \Delta J_n \left( \frac{R}{r} \right)^{n+1} P_n (cos \theta)
\end{equation}

where we have equated the sum of the particular and the homogeneous solutions for the potential interior to the body to the potential exterior to the body. For a spherical planet, we equate terms corresponding to the $n^{th}$ Legendre polynomial and set on $r = R$, which gives us:

\begin{equation}
V_{pn} (R) + A_n j_n(kR) = - \Delta J_n
\end{equation}

However, for an oblate surface, $r = R$ no longer represents the boundary and is replaced by $r = R \left( 1 + \varepsilon P_2 \left( cos \theta \right) \right)$, which introduces a $\theta$ dependence in the expression. Since $\varepsilon$ is small, we can do a Taylor series expansion of $V_{pn}$ and $j_{kr}$ around $r = R$ to obtain an approximate expression for these quantities when the planet is oblate.

\begin{eqnarray}
V_{pn} \approx V_{pn} (R) + \frac{\mathrm{d} V_{pn}}{\mathrm{d} r} \varepsilon P_2 (cos \theta)  R \\ 
j_n \approx j_n (kR) + \frac{\mathrm{d} j_n (kr)}{\mathrm{d} r} \varepsilon P_2 (cos \theta) R
\end{eqnarray}

Substituting these expressions in the boundary condition gives terms with products of two Legendre polynomials $P_2$ and $P_n$. The product of two Legendre polynomials can be written as a sum of Legendre polynomials weighted by the Clebsch-Gordon coefficients \citep{Adams1878}.

\begin{equation}
P_p (cos \theta) \; P_q (cos \theta) = \sum_{i = 0}^{q} \frac{ B_i B_{p-i} B_{q-i} }{ B_{p+ q -i} } \;  \frac{2p + 2q - 4i + 1}{2p + 2q -2i +1} \; P_{p+q-2i} (cos \theta) \; \mathrm{for} \; p \geq q
\end{equation}

where the $B$ coefficients are given by:

\begin{equation}
B_m = \frac{ \left( 1/2 \right) \left( 1/2 +1 \right) ... \left( 1/2 + m - 1 \right) }{m !} \; \mathrm{with} \; B_0 = 1
\end{equation}

Given that we equate coefficients of a particular $n^{th}$ order Legendre polynomial at the boundary, including the effect of oblateness via $P_2$ couples the $n^{th}$ gravitational moment with the order $n-2$ and the $n+2$ moments by including terms that emerge from the multiplication of $P_2$ with $P_n$. The equality of interior and exterior potential for an oblate planet therefore gives us:

\[
V_{pn} (R)  + j_n (kR)
+ \varepsilon
\begin{bmatrix}
   \frac{1}{4} \left( R V_{pn}'(R) + \pi A_n j_n'(kR)  \right) +  \\
    \frac{3}{8} \left( R V_{p(n-2)}'(R) + \pi A_{n-2}  j_{n-2}'(kR) \right)  +  \\
    \frac{3}{8} \left( R V_{p(n+2)}'(R) + \pi A_{n+2} j_{n+2}'(kR) \right) 
\end{bmatrix}
=
- \Delta J_{n,oblate}
\]

Here, the Clebsch-Gordon coefficients are exact in the limit of large $n$ but serve as a useful approximation for the range of $n$ we are interested in. The expression in the parenthesis is related to gravitational moments for a spherical planet by the continuity of the gradient of the potential and is simply given by:

\begin{equation}
R V_{pn}'(R) + \pi  A_n  j_n'(kR) = (n+1) \Delta J_n
\end{equation}

Substituting these relations into our expression for $\Delta J_{n,oblate}$, we obtain the final form of our result:

\begin{equation}
\Delta J_{n,oblate} = \Delta J_{n,sph}  - \varepsilon \bigg[ \frac{1}{4} (n+1) \Delta J_{n,sph} + \frac{3}{8} (n-1) \Delta J_{n-2,sph}  +  \frac{3}{8} (n+3) \Delta J_{n+2,sph}  \bigg] 
\end{equation}

\section{Numerical Calculations}
All the numerical calculations are carried out both in Python and Mathematica. Here, we describe the details of the procedure that we adopt for numerical calculations of $\Delta J_n$. Instead of calculating $\Delta J_n$ directly for sinc-like differential flows profiles, we write them as a sum of 15 exponential profiles (Equation~\ref{eq:exponentials}).
Numerical solution for exponentially decaying flow profiles is rapid. In addition, since we use the same 15 exponentials for all profiles and its only their coefficients that vary, we just need to perform this calculation once for each of these exponential functions. Thereafter, $\Delta J_n$ for any profile can simply be written as a sum of $\Delta J_{n,i}$ obtained from the $i^{th}$ exponential profile and weighted by the respective coefficient due to the theory's linear nature:

\begin{equation}
    \Delta J_n = \sum_{i=1}^{15} \alpha_i \Delta J_{n,i}
\end{equation}

For the exponential profiles, we then calculate $\Delta J_n$ as follows:
\begin{itemize}
    \item We integrate Equation~\ref{eq:Vrot_integrate} to find $\delta V_{rot}$ using Sympy integrate (Python) or simply `Integrate' in {\it Mathematica} \textcopyright.
    \item Given $\delta V_{rot}$, we can solve Equation~\ref{eq:V_p_diffeq} to find $V_p$ using any of the differential equation solving routines (Sympy `dsolve' or {\it Mathematica}\textcopyright `DSolve'). The calculation is analytical thus far and has already been described in detail in Appendix B.
    \item This is step where we deviate from our analytical solutions. Instead of using an approximation for the Legendre polynomials, we numerically integrate Equation~\ref{eq:V_p_projection2} to find the projection of $V_p$ onto spherical coordinates. The numerical integration is performed using `NIntegrate' in {\it Mathematica}\textcopyright or Numpy `trapz' in Python (the trapezium method only introduces small errors but is much faster and allows us to perform MCMC). In addition, Equation~\ref{eq:V_p_projection2} tells us that calculating $\mathrm{d}V_{pn} / \mathrm{d}r$ simply involves differentiating $V_p$ with respect to $r$ inside the integral (which is an analytical step), followed by numerical integration of the expression. We can then use $V_{pn}$ and $V'_{pn}$ to calculate $\Delta J_n$ from Equation~\ref{eq:deltaj_n}.
    
\end{itemize}

\end{document}